# KiloDA: Reconstructing kilometer-scale near-surface wind states from sparse station observations

Yihan Zheng[1,3], Ya Wang*[2,3], Gang Huang*[2,3], Shenming Fu[2,3], Haijie Li[2,3] and Yujiang Cai[2,3]

[1] *Center for Monsoon System Research, Institute of Atmospheric Physics, Chinese Academy of Sciences, Beijing 100029, China.*

[2] *China Key Laboratory of Earth System Numerical Modeling and Application, Institute of Atmospheric Physics, Chinese Academy of Sciences, Beijing 100029, China.*

[3] *University of Chinese Academy of Sciences, Beijing 100049, China.*

Corresponding authors: Dr. Ya Wang (wangya@mail.iap.ac.cn)

Dr. Gang Huang (hg@mail.iap.ac.cn)

# Abstract

Accurate kilometer-scale near-surface winds are important for understanding atmospheric processes over complex terrain, yet remain difficult to reconstruct from sparse and unevenly distributed observations. Here we introduce KiloDA, a diffusion framework for hourly kilometer-scale wind reconstruction from surface stations. KiloDA learns the statistical distribution and spatial structure of wind fields from historical 3-km Weather Research and Forecasting (WRF) model forecasts. At each reconstruction time, no contemporaneous WRF field is used. Instead, station observations provide the only constraints on the current atmospheric state and guide posterior sampling from the learned prior. In idealized WRF experiments, KiloDA recovers localized wind structures when only 0.24% of grid cells are observed and shows an overall advantage over conventional interpolation across terrain conditions and wind speed regimes. This capability largely transfers to real observations. In a fully withheld region, KiloDA reduces the median wind speed root mean square error (RMSE) by 19% relative to ERA5 reanalysis, using only observations outside the region, with the largest improvements over high-elevation and high-relief terrain. A random station holdout further confirms that this advantage extends across different complex-terrain locations and holdout configurations. These results show that historical model archives can provide useful structural knowledge for reconstructing kilometer-scale wind fields from sparse observations without requiring an accurate model estimate of the current atmospheric state.

# Introduction

Kilometer-scale wind fields provide essential information for wind resource assessment, extreme weather analysis and local wind forecasting[1–3]. Over complex terrain, wind speed and direction can vary sharply over short distances because of variations in surface elevation, land surface heterogeneity and terrain effects such as flow channeling, ridge acceleration and sheltering on the lee side[4–7]. Yet surface stations sample this highly heterogeneous field only at sparse and unevenly distributed locations, with particularly limited coverage in mountainous regions[8,9]. The central challenge is therefore to infer continuous kilometer-scale wind states from sparse observations while retaining the fine-scale structures created by complex terrain.

Spatial interpolation methods such as inverse distance weighting (IDW) and kriging provide conventional approaches for filling gaps between irregularly distributed stations[10,11]. However, because they rely mainly on distance or spatial covariance, they have limited ability to represent terrain-dependent wind structures that are not directly sampled by observations. Their performance degrades with increasing terrain complexity and spatial variability, often resulting in oversmoothed fields and attenuated extreme values[12–15]. Data assimilation provides a more systematic way to integrate observations with model information[16]. Global reanalysis products such as the fifth-generation atmospheric reanalysis from the European Centre for Medium-Range Weather Forecasts (ERA5) offer spatially complete and physically consistent atmospheric states, but their relatively coarse resolution limits the representation of terrain-induced wind structures over complex topography[17–19]. Kilometer-scale regional models such as the Weather Research and Forecasting (WRF) model can better resolve terrain variability and local wind structures and can also assimilate observations to improve the atmospheric state[20–22]. However, continuously running kilometer-scale numerical models and assimilating observations into them is computationally expensive, limiting their use for efficient wind analysis over long periods at high spatial resolution[23,24].

Deep learning has rapidly advanced weather and climate research, with data-driven systems such as Pangu-Weather[25] and FuXi[26] demonstrating strong performance in global weather prediction, while related diagnostic frameworks have also been used to investigate atmospheric processes[27]. Within this broader development, generative models offer a particularly promising approach for reconstructing high-dimensional atmospheric fields. Diffusion models can learn the spatial structures and statistical distribution of complete atmospheric states from historical simulations at high resolution and subsequently generate physically plausible fields without rerunning the numerical model[28–30]. Observation-guided diffusion further allows sparse measurements to constrain a pretrained generative prior during inference, providing a flexible framework for reconstructing high-dimensional states from incomplete observations[31,32]. This approach has recently been explored for atmospheric and oceanic reconstruction[33–35], but its capability at kilometer scales over complex terrain remains poorly established. In particular, it remains unclear whether extremely sparse station observations are sufficient to recover terrain-dependent wind structures in unobserved locations, and whether such capability transfers from an idealized model setting to a real observational setting.

Here, we introduce KiloDA, a kilometer-scale diffusion assimilation framework for reconstructing near-surface wind fields from sparse station observations. KiloDA learns a kilometer-scale wind field prior for the 10-m zonal (U10) and meridional (V10) wind components from historical 3-km WRF samples and uses station observations to guide reverse diffusion sampling during inference, thereby reconstructing complete wind field states under sparse observational constraints. Idealized WRF experiments show that KiloDA can reconstruct unobserved fine-scale wind structures from extremely sparse observations and consistently outperforms conventional interpolation across terrain conditions and wind speed regimes. Real-observation experiments further show that this capability transfers to complex terrain, reducing wind speed errors relative to ERA5 and interpolation methods. These results demonstrate that sparse observations combined with a learned structural prior can recover kilometer-scale wind states beyond observed locations.

# Results

## KiloDA reconstructs kilometer-scale near-surface winds from sparse stations

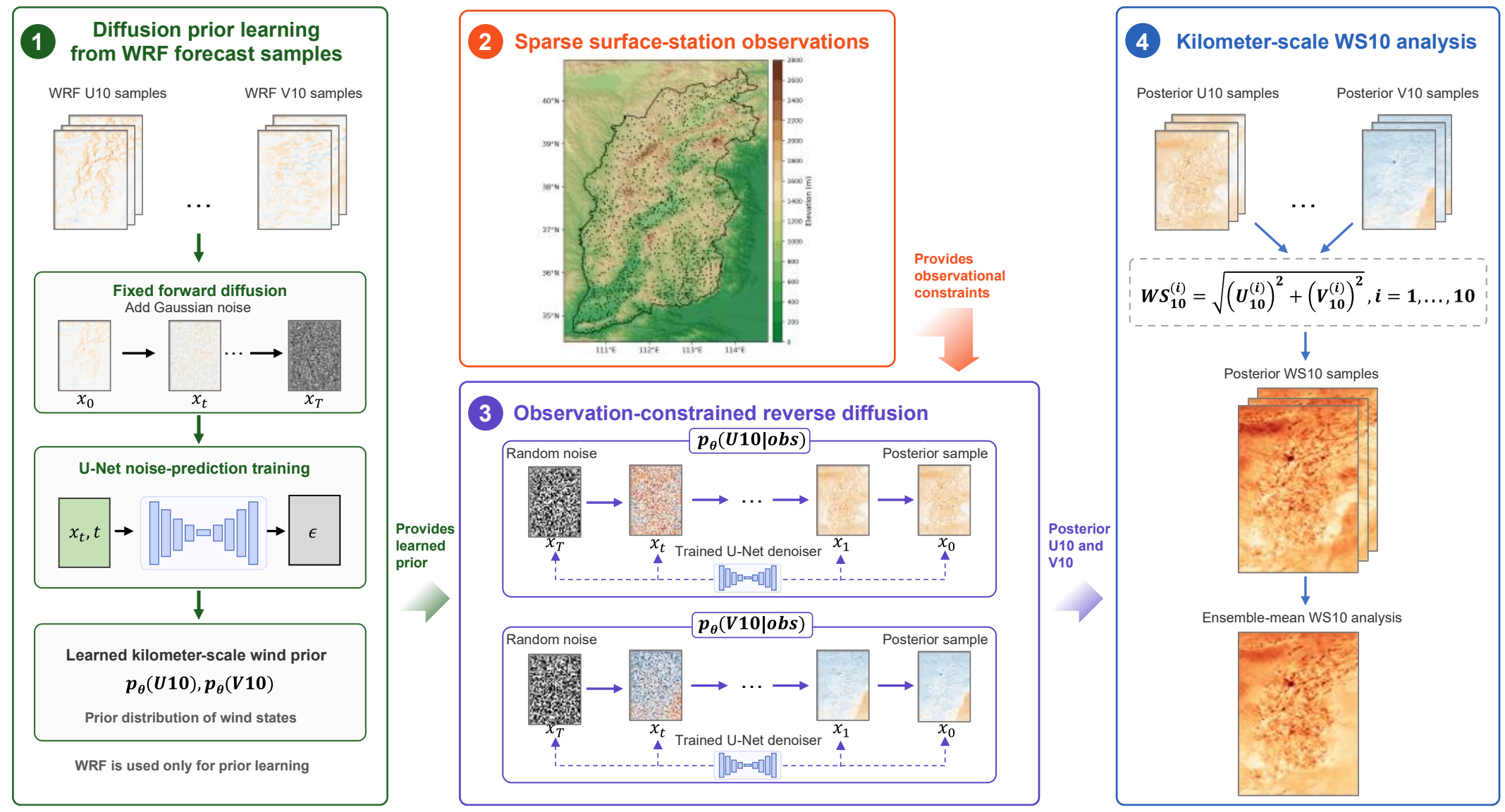

**Fig. 1 KiloDA framework.** KiloDA learns kilometer-scale wind priors from 3-km WRF model data and constrains reverse diffusion sampling with sparse surface station observations to generate continuous fields of the 10-m zonal (U10) and meridional (V10) wind components, and derive 10-m wind speed (WS10).

Figure 1 summarizes the KiloDA framework (see Methods). KiloDA first uses historical 3-km WRF forecasts to learn a generative prior for kilometer-scale wind structures. During reconstruction, it does not use the WRF forecast for that time. Instead, current station observations steer Bayesian posterior sampling to generate complete U10 and V10 fields, from which 10-m wind speed (WS10) is derived. Each reconstructed field must satisfy two requirements: it should follow the statistical distribution and spatial structure captured by the WRF prior, and agree with the available observations. In this sense, WRF defines a plausible structural manifold for kilometer-scale winds, while observations identify the current atmospheric state. As the prior is built from WRF forecasts, some forecast errors are inevitably inherited. WRF indeed shows larger station-wise root mean square error (RMSE) than ERA5, but it better captures the observed spatial variability and retains much richer fine-scale structures (Supplementary Fig. 1). This property makes WRF suitable for constructing the KiloDA prior, which relies more on realistic distributions and spatial relationships than on accurate state estimates at individual times.

## Learning a kilometer-scale wind prior from WRF forecasts

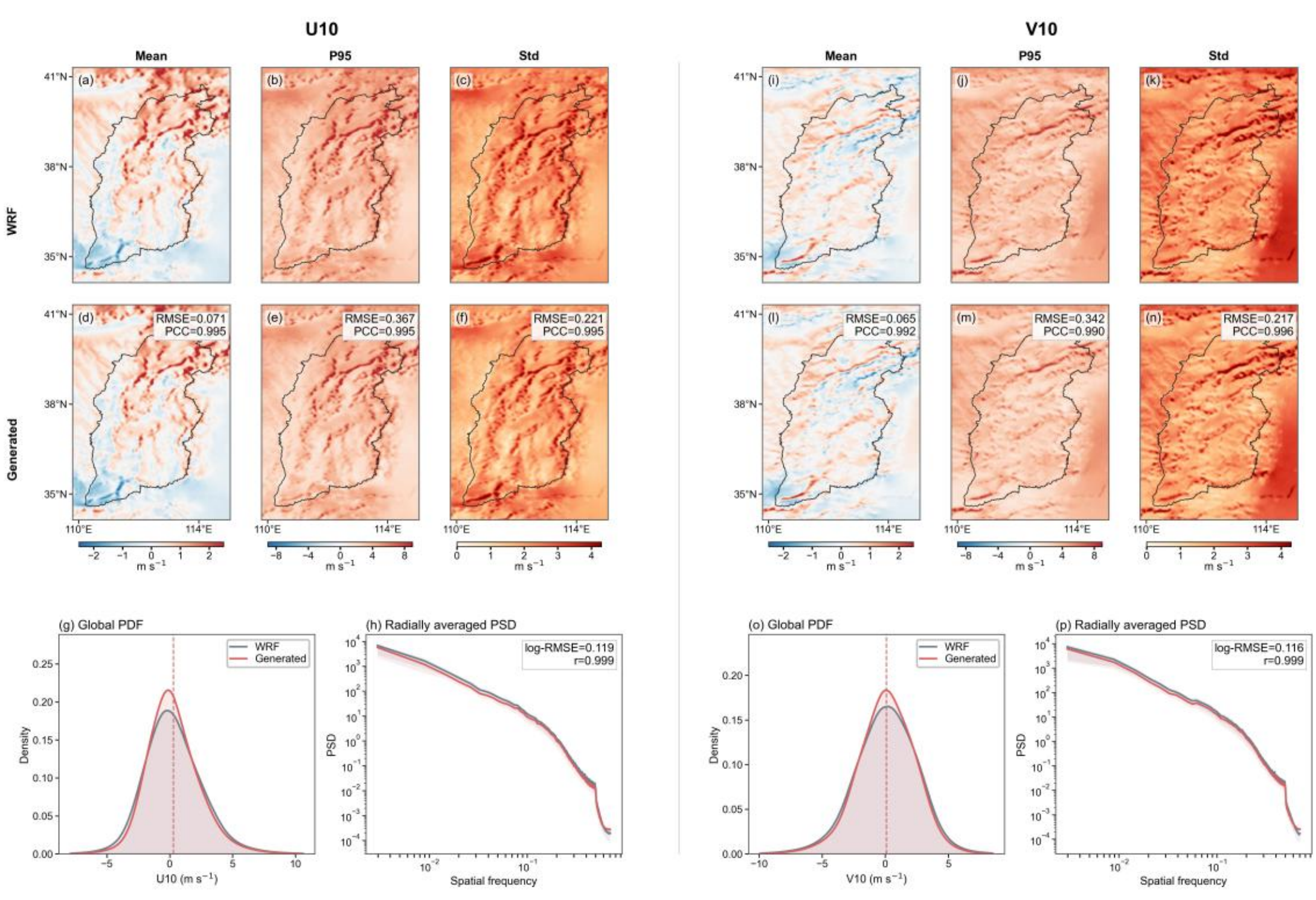

**Fig. 2 Learning a kilometer-scale wind prior from WRF forecasts.** For U10 (a–h) and V10 (i–p), the top and middle rows show the mean, 95th percentile and standard deviation fields computed from the WRF samples and generated samples, respectively. Panels (g, o) show the pooled probability density functions (PDFs), and panels (h, p) show the radially averaged power spectral density (PSD). Root mean square error

(RMSE) and pattern correlation coefficient (PCC) quantify the spatial discrepancy and pattern agreement between the generated and corresponding WRF statistical fields. In panels (h, p), log-RMSE denotes the RMSE between the log10-transformed generated and WRF mean PSDs, while the Pearson correlation coefficient ($r$) quantifies their agreement. Shading around the PSD curves denotes the interquartile range across samples. U10 and V10 are in m $s^{-1}$.

Reliable observation-constrained reconstruction requires a prior that represents both the spatial structure and intrinsic variability of kilometer-scale wind fields. We therefore first assessed whether the diffusion models could learn these characteristics from the 2024 hourly WRF wind fields. We compared 5,000 generated samples with all available WRF samples from 2024 using the mean, 95th percentile and standard deviation fields, probability density functions (PDFs) and radially averaged power spectral densities (PSDs) (Fig. 2).

The generated samples closely reproduce the WRF statistics for both U10 and V10. The mean, 95th percentile and standard deviation fields show RMSE values of 0.065–0.367 m $s^{-1}$ and pattern correlation coefficient (PCC) values of 0.990–0.996, while the probability distributions are highly consistent with those of the WRF samples. The radially averaged PSDs are also nearly identical, with Pearson correlation coefficients of 0.999 for both wind components. These results demonstrate that the diffusion models successfully learn the statistical distribution, spatial heterogeneity and multiscale structure of the kilometer-scale WRF wind fields, providing a suitable generative prior for subsequent observation-constrained reconstruction.

## Idealized reconstruction in the WRF model world

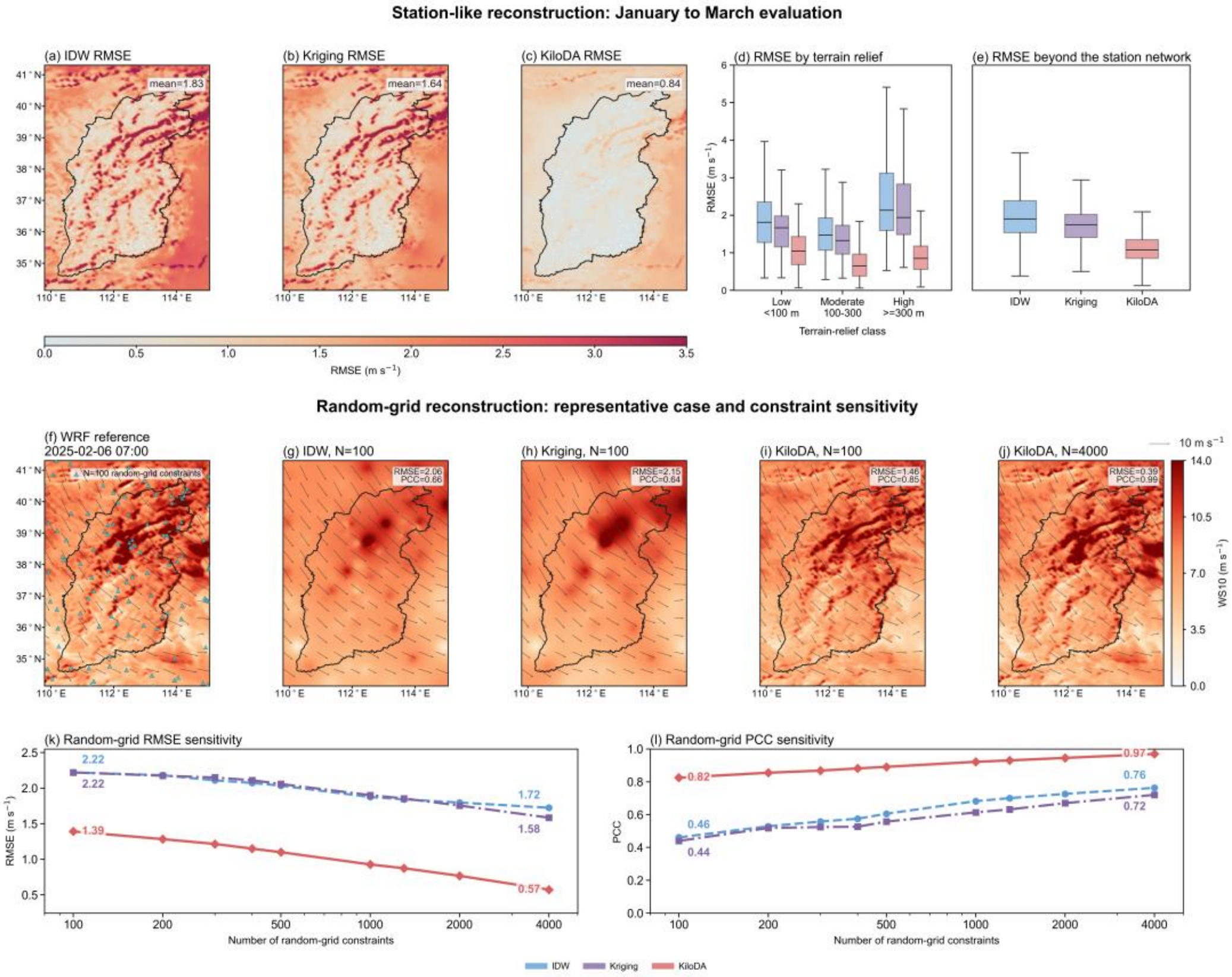

**Fig. 3 Idealized reconstruction of unobserved kilometer-scale wind states in the WRF model world.** All evaluation metrics are calculated only at unconstrained grid cells. (a–c) Grid-cell RMSE for inverse distance weighting (IDW), Kriging and KiloDA in the station-like reconstruction, evaluated over 2,117 hourly WRF reference fields from January to March 2025. Inset values indicate spatial mean RMSE. (d, e) RMSE distributions for terrain-relief classes and grid cells beyond the station network, respectively. Boxes show the interquartile range, center lines indicate medians, and whiskers extend to the most extreme values within 1.5 times the interquartile range; outliers are not shown. (f–j) Representative strong-wind case at 07:00 on 6 February 2025 in the random-grid reconstruction, showing the WRF reference with N = 100 constraint grid cells (f), IDW with N = 100 (g), Kriging with N = 100 (h), KiloDA with N = 100 (i), and KiloDA with N = 4,000 (j). Inset values in (g–j) indicate RMSE and PCC. (k, l) Mean RMSE and PCC as functions of the number of randomly selected constraint grid cells during the 32-hour strong-wind period. Shading in (a–c) denotes RMSE, while shading in (f–j) denotes WS10 and arrows denote horizontal wind vectors. Cyan markers in (f) indicate constraint grid cells. WS10 and RMSE are in m $s^{-1}$.

To isolate the intrinsic reconstruction capability of KiloDA from real-world observation errors and data mismatches, we first evaluated it in an idealized setting where the complete wind field is known and sparse observations are sampled directly from the same model world. Using independent WRF fields from January to March 2025 (JFM) as complete reference states, we conducted idealized reconstruction experiments in which pseudo-observations were extracted from selected grid cells and used as observational constraints for KiloDA and as inputs to the interpolation baselines.

In the station-like experiment, IDW and Kriging show clear spatially organized errors rather than uniformly distributed errors (Fig. 3a, b). Large errors form elongated bands across regions of strong terrain variation, indicating that conventional interpolation has difficulty recovering wind structures that vary rapidly with topography at locations without direct observational constraints. In contrast, KiloDA substantially suppresses these terrain-related error bands and produces much lower errors across most of the domain (Fig. 3c). This terrain dependence becomes clearer as terrain relief increases. Errors from IDW and Kriging rise markedly with terrain relief, whereas the increase is much smaller for KiloDA (Fig. 3d). Outside the station network, where grid cells are relatively far from the observational constraints, interpolation errors also remain high. KiloDA maintains substantially lower errors in these areas, reflecting a stronger ability to reconstruct wind fields beyond the spatial coverage of the observational constraints (Fig. 3e). Across all 2,117 hourly fields, the spatial mean RMSE is reduced from 1.83 m $s^{-1}$ for IDW and 1.64 m $s^{-1}$ for Kriging to 0.84 m $s^{-1}$ for KiloDA.

We next examined a representative strong-wind event to determine whether sparse observations can recover localized wind structures rather than only the broad background state. The WRF reference shows a narrow and highly heterogeneous strong-wind band extending across the central and northern part of the domain, with multiple local maxima and pronounced spatial variations, including a distinct maximum along the eastern periphery of the domain (Fig. 3f). With only 100 constraint grid cells, IDW and Kriging recover only a broad strong-wind center and smooth the reference field into a relatively simple pattern, missing most of the narrow terrain-related structures and the distinct maximum along the eastern periphery (Fig. 3g, h). Their wind directions are also much more spatially uniform than in the reference field. In contrast, KiloDA recovers the main elongated strong-wind band, much of its internal spatial variability and the associated changes in wind direction even with the same 100 constraint grid cells (Fig. 3i). Increasing the number of constraint grid cells to 4,000 further refines these structures, with the local maximum along the eastern periphery well reproduced and the overall field closely matching the WRF reference (Fig. 3j).

This behavior is consistent across constraint densities. Even with only 100 constraint grid cells, or 0.24% of the domain, KiloDA achieves an RMSE of 1.39 m $s^{-1}$ and a PCC of 0.82, compared with RMSE values above 2 m $s^{-1}$ and PCC values below 0.5 for the interpolation methods. As more constraint grid cells are added, KiloDA progressively approaches the reference field, with RMSE decreasing to 0.57 m $s^{-1}$ and PCC increasing to 0.97 with 4,000 constraint grid cells (Fig. 3k, l). This advantage is maintained across terrain-relief and elevation classes under different numbers of constraint grid cells (Supplementary Fig. 3). Notably, the RMSE gap between KiloDA and the interpolation methods does not diminish as the number of constraint grid cells increases, indicating that increasing the number of constraints alone is insufficient to eliminate the limitations of conventional interpolation in reconstructing unobserved spatial structures. These results show that the main advantage of KiloDA is not simply lower pointwise error. By drawing on the learned kilometer-scale wind prior, it can recover terrain-related spatial structures that are not directly resolved by the sparse observations themselves.

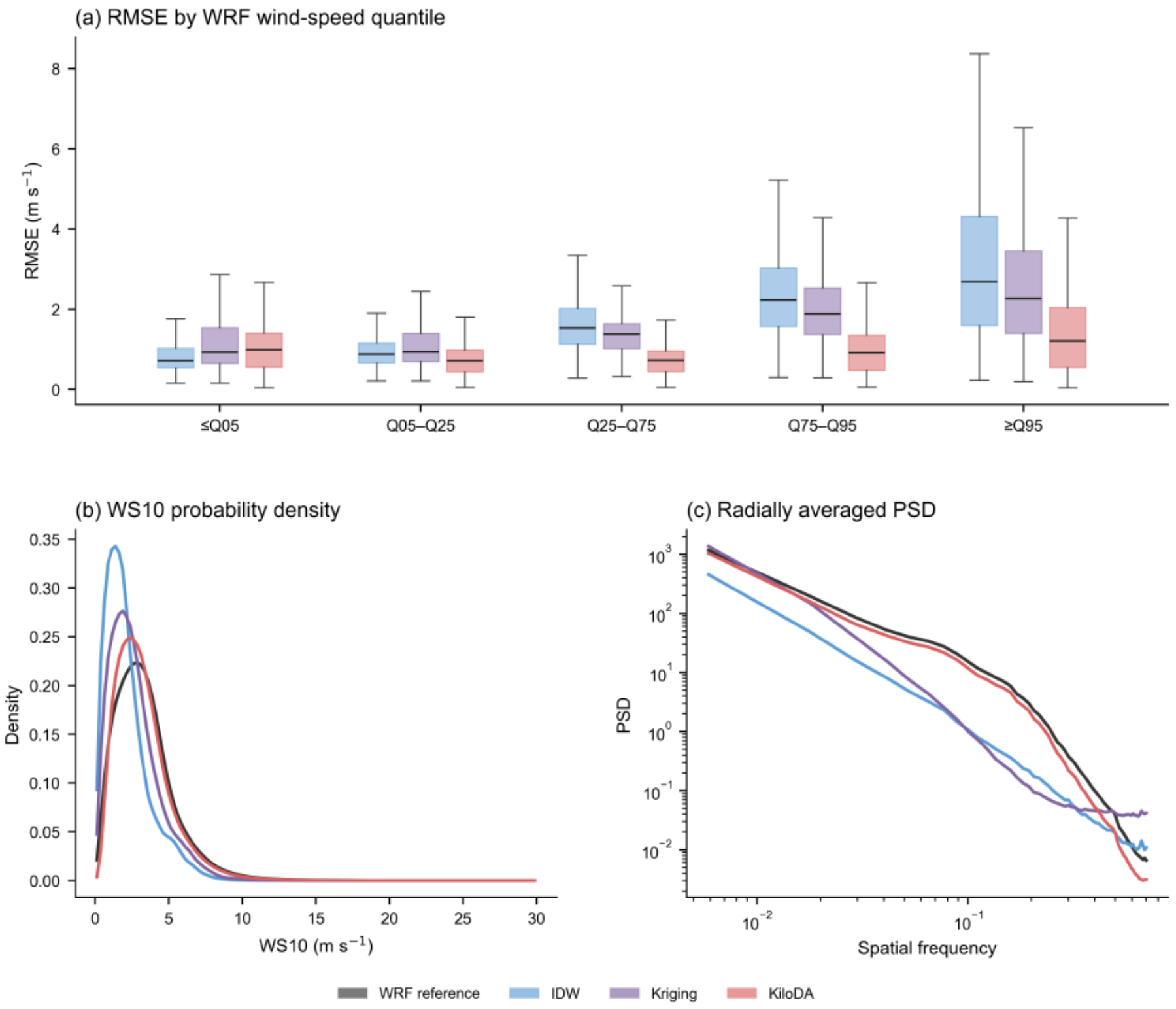

**Fig. 4 Reconstruction errors across WRF wind speed regimes and characteristics of the reconstructed fields.** (a) Distributions of WS10 RMSE for IDW, Kriging and KiloDA over unconstrained grid cells, stratified by WRF reference wind speed quantiles (≤Q05, Q05–Q25, Q25–Q75, Q75–Q95 and ≥Q95) during the January to March 2025 evaluation period. Quantile thresholds are defined separately at each grid cell from its WRF wind speed time series during the evaluation period. Boxes show the interquartile range, center lines indicate medians, and whiskers extend to the most extreme values within 1.5 times the interquartile range; outliers are not shown. (b) Probability density functions of WS10 over the full field for the WRF reference, IDW, Kriging and KiloDA. (c) Radially averaged PSD of WS10 for the WRF reference, IDW, Kriging and KiloDA, averaged over sampled wind fields. WS10 and RMSE are in m $s^{-1}$.

To further examine what types of wind fields can be reconstructed, we evaluated the station-like experiment across different wind speed regimes and compared the statistical and spatial characteristics of the reconstructed fields (Fig. 4). The advantage of KiloDA becomes increasingly clear as wind speed increases. At weak winds below Q05, the three methods show comparable errors and IDW performs slightly better. From Q05 upward, however, KiloDA consistently produces lower RMSE, and the separation from IDW and Kriging grows substantially in the upper wind speed regimes. For winds above Q75, the median RMSE of KiloDA is roughly half that of the interpolation methods, indicating that the learned prior is particularly important when reconstructing stronger and more spatially heterogeneous winds (Fig. 4a).

The reconstructed wind speed distributions further show why the interpolation methods lose skill (Fig. 4b). IDW produces a much narrower distribution with excessive probability at weak wind speeds, while Kriging shows a similar but weaker tendency. Both methods therefore suppress the variability and stronger winds present in the WRF reference. In contrast, the KiloDA distribution closely follows the WRF reference across the main body of the distribution and extends into a similar high wind speed tail. The same contrast appears across spatial scales (Fig. 4c). IDW and Kriging rapidly lose spectral power from intermediate to small spatial scales, consistent with their overly smooth reconstructed fields. KiloDA retains substantially more spectral energy and closely follows the WRF reference across most resolved scales. Together, these results show that KiloDA does more than reduce pointwise error. It preserves the wind speed variability and kilometer-scale spatial structure that conventional interpolation tends to smooth out, particularly under stronger wind conditions.

## Reconstruction from real station observations

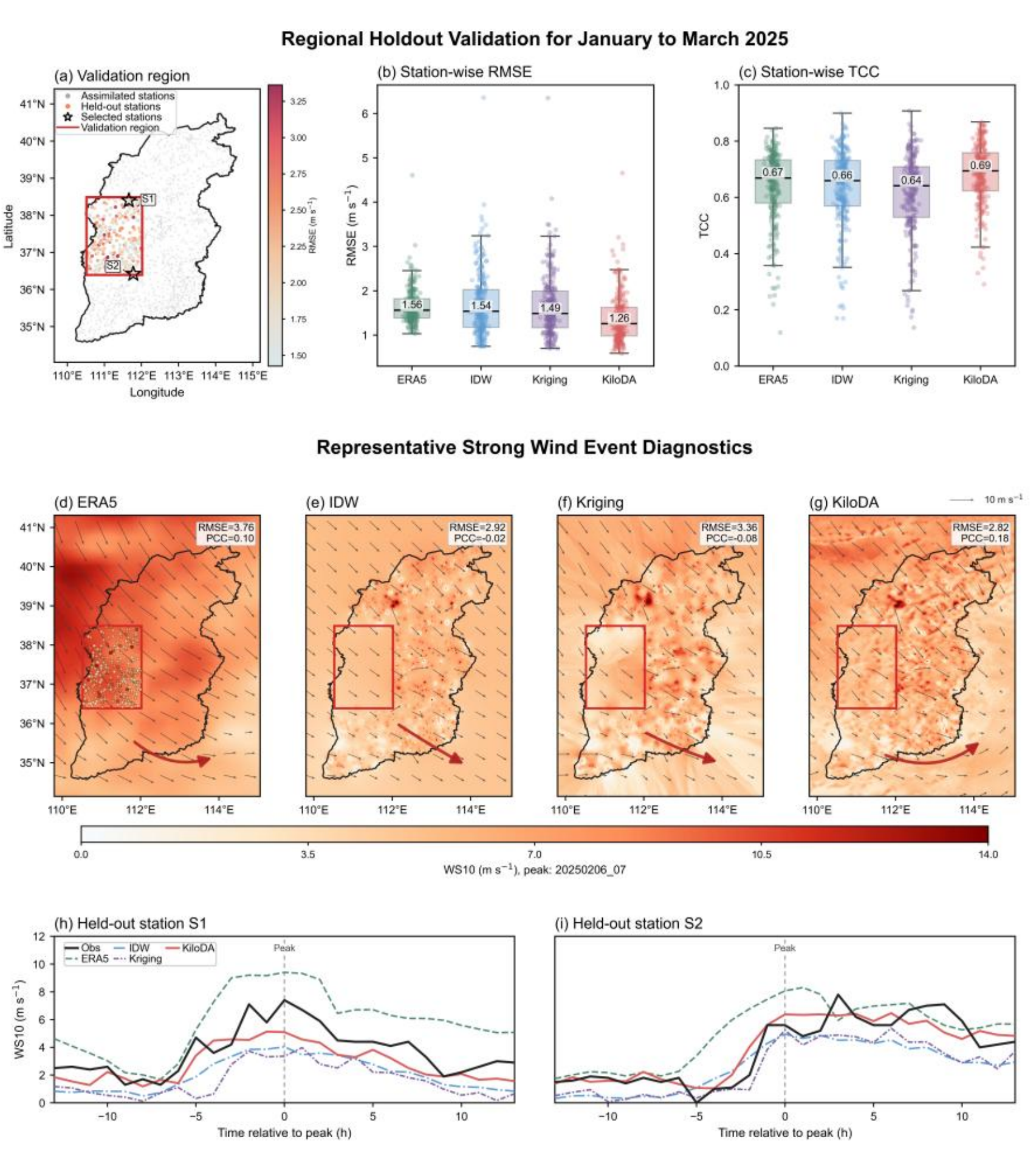

**Fig. 5 Validation results for a fully withheld region using real observations.** (a) Validation setup for January to March 2025. All stations within the red rectangle were withheld from the reconstruction inputs and used only for independent validation, while stations outside the region provided observational constraints. S1 and S2 mark two representative withheld stations. (b, c) Station-wise WS10 RMSE and temporal correlation coefficient (TCC) for ERA5, IDW, Kriging and KiloDA at the withheld stations. Boxes show the interquartile range, center lines indicate medians, and whiskers extend to the most extreme values within 1.5 times the interquartile range. Points denote individual withheld stations. (d–g) WS10 fields and wind vectors from ERA5, IDW, Kriging and KiloDA at the peak of the representative strong wind event (07:00 on 6 February 2025); inset values show RMSE and the spatial PCC across the withheld stations. Red arrows highlight the cyclonic circulation pattern. (h, i) WS10 time series around the event peak at S1 and S2. WS10 and RMSE are in m $s^{-1}$.

The idealized experiments established that KiloDA can reconstruct unobserved kilometer-scale wind states when the prior and observations are statistically consistent and the complete reference field is known. The more important question is whether this reconstruction capability can transfer to real station observations, where observations are unevenly distributed and may differ from the model-based prior in both sampling characteristics and representativeness. We therefore performed a contiguous withheld-region validation in which all observations inside a target region were removed from the reconstruction inputs, allowing us to test whether observations outside the region could constrain the wind state within a completely unobserved area (Fig. 5). The contiguous target region, located mainly in the Lüliang mountainous area of western-central Shanxi, was selected based on the spatial pattern of ERA5 errors during JFM 2025, as described in Methods. All 238 stations inside the region were withheld and used only for independent evaluation, while the remaining 1,102 stations provided identical observational constraints to IDW, Kriging and KiloDA.

Over the evaluation period, KiloDA achieves a median WS10 RMSE of 1.26 m $s^{-1}$ across the withheld stations, corresponding to reductions of 19.2%, 18.2% and 15.4% relative to ERA5, IDW and Kriging, respectively (Fig. 5b). This advantage is not confined to a small subset of stations. KiloDA yields lower RMSE than ERA5, IDW and Kriging at 66.4%, 88.2% and 89.5% of the withheld stations, respectively, with the improvements broadly distributed across the target region (Supplementary Fig. 4). Notably, this pattern persists across terrain-relief and elevation groups, with the largest improvements occurring in the highest-relief and highest-elevation categories (Supplementary Fig. 5). A similar advantage is evident in temporal agreement, with KiloDA achieving the highest median temporal correlation coefficient (TCC) of 0.69 across the withheld stations (Fig. 5c).

A representative strong wind event further illustrates differences in the spatial organization of the reconstructed wind fields (Fig. 5d–g). For IDW and Kriging, local wind speed variations appear mainly as discontinuous patches around the constraint stations, whereas the fully withheld region is notably smooth and shows little internal spatial variability. Similar smoothing is also evident in other unconstrained areas. In contrast, KiloDA exhibits greater internal spatial variability and more heterogeneous wind speed patterns within the fully withheld region, with spatial variation also evident farther from the constraint stations. The wind vectors further reveal differences in the reconstructed flow structure. ERA5 exhibits a coherent cyclonic circulation pattern over the southern and southeastern parts of the study domain, as highlighted by the red arrows, whereas the interpolation methods produce much more uniform flow directions (Fig. 5d–f). KiloDA, however, reconstructs the cyclonic circulation pattern seen in ERA5, suggesting that it captures coherent large-scale flow features (Fig. 5g). Consistent with the spatial comparison, the time series at withheld stations S1 and S2 show that KiloDA more closely follows the observed wind speed evolution during the strong wind event (Fig. 5h, i).

To test whether the reconstruction advantage was specific to the contiguous target region, we further conducted a random holdout experiment by withholding 100 stations selected from complex-terrain areas (Supplementary Fig. 6). Despite the different spatial configuration of the withheld stations, KiloDA consistently achieves lower RMSE than all baselines while matching or exceeding them in TCC over the three-month evaluation period. These results indicate that the reconstruction advantage extends across different complex terrain locations and does not depend on the particular spatial configuration of the contiguous target region.

Together, these results show that KiloDA can reconstruct wind fields within a contiguous region with no internal observational constraints using sparse observations outside the target region. Its overall advantage is maintained across different terrain conditions and different holdout configurations, demonstrating that the reconstruction capability established in the idealized experiments largely transfers to real observations over complex terrain. However, the improvement is not uniform across the wind speed distribution. KiloDA remains competitive through most of the observed distribution but shows slightly higher RMSE than ERA5 in the upper 5% of wind speeds (Supplementary Fig. 5).

## Regional influence of sparse observations

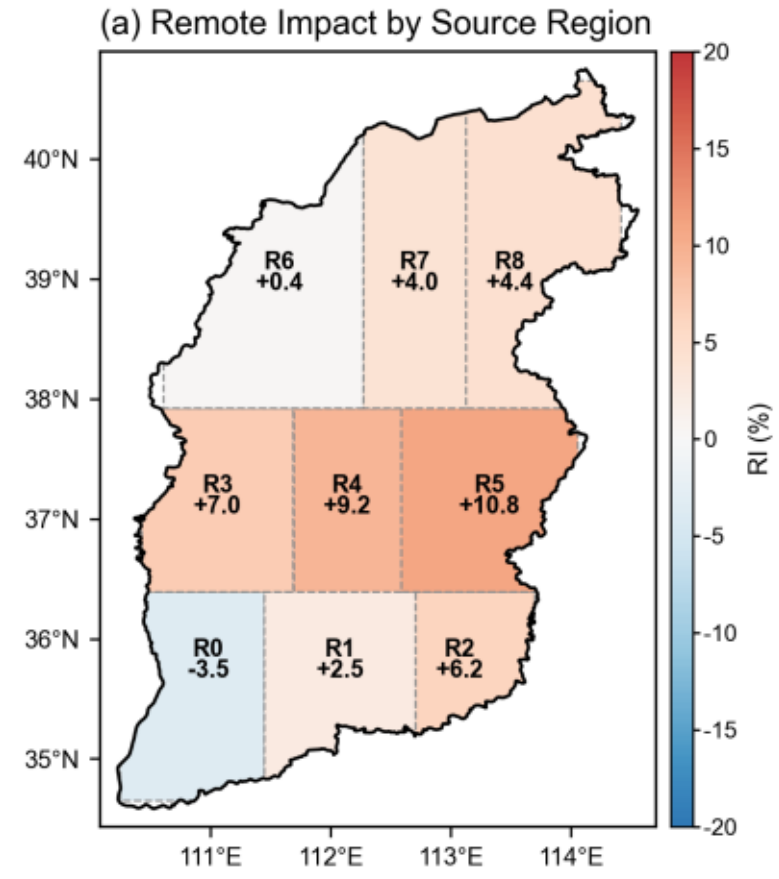

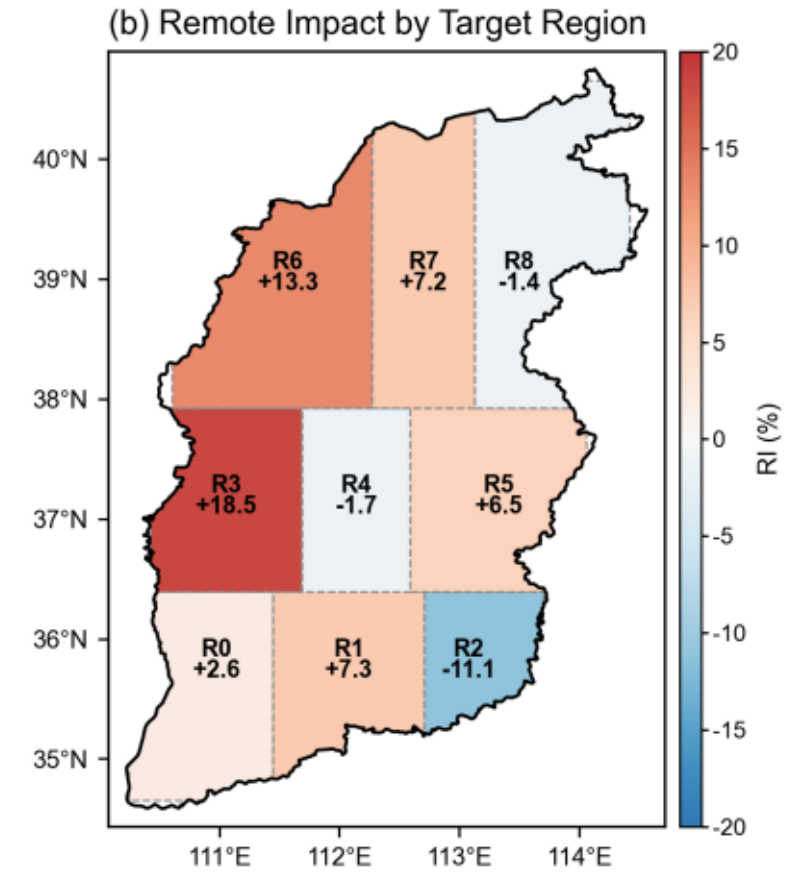

**Fig. 6 Regional remote impact (RI) on KiloDA reconstruction performance.** (a) Mean RI for each source region, averaged across all other target regions. (b) Mean RI for each target region, averaged across all other source regions. Positive RI values indicate that KiloDA produces a lower target region WS10 RMSE than ERA5.

To examine how KiloDA's remote reconstruction capability varies with the spatial location of the observational constraints, we conducted regional influence experiments across nine subregions (R0–R8) during the same representative strong wind event (Fig. 6).

The influence of regional observations on remote reconstruction shows clear spatial heterogeneity. Observations from R5 and R4 produce the largest mean improvements in remote reconstruction, with mean relative RMSE reductions of 10.8% and 9.2%, respectively, across the other target regions. By contrast, R0 shows a small mean RMSE increase of 3.5% relative to ERA5 across the other target regions, suggesting that remote improvements are not consistently positive across source regions. Remote reconstruction also varies among target regions, with mean relative RMSE reductions reaching 18.5% for R3 and 13.3% for R6, whereas R2 shows a mean RMSE increase relative to ERA5.

As an additional diagnostic, the regional influence experiments extend the single withheld region test by considering multiple pairings of source and target regions across the study domain. They show that the remote influence of observations is spatially heterogeneous and depends on the specific pairing of source and target regions.

## Discussion

In this study, we develop KiloDA, a kilometer-scale diffusion assimilation framework for reconstructing near-surface wind fields from sparse station observations. Unlike conventional interpolation, which estimates unobserved values directly from the available observations, KiloDA learns a generative prior for kilometer-scale wind fields from historical high-resolution WRF fields and uses current station observations to constrain the reconstruction of spatially complete wind fields. This distinction is particularly important over complex terrain, where kilometer-scale winds exhibit strong spatial heterogeneity, local gradients, and terrain-modulated structures that are difficult to recover using conventional interpolation from sparse observations.

The performance implications of this distinction are evident in both the idealized and real-observation experiments. In the idealized WRF experiments, KiloDA reconstructs unobserved kilometer-scale wind structures under very sparse observational constraints and maintains an overall advantage over the interpolation methods across different observation densities, terrain conditions and wind speed regimes. More importantly, this reconstruction capability is largely retained when KiloDA is conditioned on real station observations. Even when all stations within a contiguous region are withheld, KiloDA achieves lower errors than ERA5 and the interpolation baselines at most withheld stations, with this advantage persisting across different terrain conditions and holdout configurations.

Further, the regional influence experiments provide a complementary diagnostic of how observational information propagates beyond observed regions. The heterogeneous responses among different source and target regions indicate that remote observational influence may not be determined by distance alone, but a detailed interpretation of the value of the observation network is beyond the scope of the present study.

An important limitation is that the current evaluation focuses mainly on Shanxi and its surrounding complex-terrain region during a winter-to-spring period, and the applicability of KiloDA to other seasons, terrain-climate regimes and weather processes requires further validation. Future work should extend the framework to longer periods, multiple regions and broader weather backgrounds, while explicitly addressing the adaptation of grid-scale generative priors to point observations. Overall, KiloDA provides a practical framework for kilometer-scale wind reconstruction from sparse observations, while the reduced performance at the upper tail of the real-observation wind-speed distribution highlights an important limitation for future work.

# Methods

## WRF simulations for prior training

KiloDA uses diffusion models to learn generative priors of kilometer-scale near-surface wind fields from historical WRF[20] simulations. These simulations are not assumed to represent the true atmospheric state. Instead, they provide dynamically plausible and spatially continuous samples from which the diffusion models learn the spatial organization, local gradients and statistical variability of wind fields over complex terrain.

The WRF simulations used Global Forecast System (GFS) [36] fields to provide the initial and lateral boundary conditions for the outer domain and employed a two-domain, one-way nested configuration, with horizontal grid spacings of 15 and 3 km for the outer and inner domains, respectively. The model used 37 vertical levels. The outer domain received external lateral boundary conditions at 3-h intervals, while the inner domain was laterally forced by the outer domain. The model was nonhydrostatic and used adaptive time stepping. The main physics configuration included Thompson microphysics, RRTMG longwave and shortwave radiation, the revised MM5 Monin–Obukhov surface layer scheme, the Noah land surface model and the Yonsei University planetary boundary layer scheme. The Kain–Fritsch cumulus scheme was used in the outer domain but not in the 3-km inner domain. Terrain-related options including slope radiation, topographic shading and topographic wind correction were enabled.

Observation nudging was applied to wind, temperature and moisture with internal quality control.

The training dataset consisted of hourly U10 and V10 fields from the 3-km inner domain during 2024. These fields were cropped to a common 240 × 176 grid covering Shanxi Province and the surrounding areas. Before training, each wind component was linearly scaled to [−1, 1] using component-specific minimum and maximum values derived from the training dataset. After sampling, the reconstructed components were transformed back to physical units. For all gridded wind fields used in subsequent analyses, the corresponding WS10 was calculated as

$$WS10 = \sqrt{U10^2 + V10^2} \quad (1)$$

## Station observations

Surface meteorological observations from the China Meteorological Administration were used to constrain KiloDA reconstructions and served as the primary basis for validation in the real-observation experiments. The station archive included hourly 10-m wind speed and wind direction together with station coordinates for 2024 and JFM 2025. The 2024 observations were used only for the diagnostic comparison shown in Supplementary Fig. 1, whereas the JFM 2025 observations were used for observational conditioning and real-observation validation. Wind speed or wind direction values coded as 999999 were treated as missing and excluded. Wind speed values outside 0–50 m $s^{-1}$ and wind direction values outside 0–360° were also excluded, except for calm-wind records coded as 999017, corresponding to wind speeds ≤0.2 m $s^{-1}$. These calm-wind records were retained during preprocessing. Stations with fewer than 100 valid records or clear temporal-quality anomalies were also excluded. To ensure consistency with the WRF variables and the KiloDA reconstruction targets, station wind speed and direction were converted into U10 and V10 wind components. Each station was matched to the nearest WRF model grid cell according to its geographic coordinates, thereby establishing a correspondence between station observations and the kilometer-scale model grid. When multiple stations were assigned to the same grid cell, their U10 and V10 components were averaged. After quality control and grid matching, the retained observations covered 1,340 grid cells. Because hourly observations were intermittently missing, the number of valid observational constraints varied over time, with an average of approximately 1,294 grid cells per hour.

## ERA5 and interpolation baselines

ERA5 reanalysis[17] was used as an external benchmark representing a widely used coarse-resolution gridded product. Hourly ERA5 U10 and V10 fields were obtained for 2024 and JFM 2025 at their native horizontal resolution of 0.25° × 0.25° and bilinearly interpolated to the 240 × 176 analysis grid. The 2024 fields were used only for the diagnostic comparison in Supplementary Fig. 1, while the JFM 2025 fields were used as the external benchmark in the real-observation experiments. ERA5 was not used as an input, background field, training target or sampling constraint in KiloDA.

IDW[37] and Ordinary Kriging[38] were additionally used as conventional spatial interpolation baselines. Both methods were applied separately to U10 and V10 at each analysis time, from which WS10 was derived. For IDW, each wind component $X \in \{U10, V10\}$ at target location $x$ was estimated as

$$\hat{X}(x,t) = \frac{\sum_{i \in S_t} w_{i(x)} x_{i(t)}}{\sum_{i \in S_t} w_{i(x)}},\ w_{i(x)} = d_{i(x)}^{-2} \tag{2}$$

where $\mathcal{S}_t$ denotes the set of valid constraint stations at time $t$, and $d_{i(x)}$ is the Euclidean distance between the target and constraint locations. When a target location coincided with a constraint location, the observed value at that location was used directly. Ordinary Kriging used an exponential semivariogram with six lag bins, with parameters estimated automatically at each analysis time and predictions based on the nearest 32 valid constraint stations. IDW, Kriging and KiloDA used the same valid observational constraints in each experiment.

## KiloDA model

KiloDA is a reconstruction framework that combines diffusion wind-field priors with sparse observational constraints. It first learns the spatial structure and statistical distribution of near-surface winds from kilometer-scale WRF fields. During reverse diffusion, station-observation consistency is imposed to generate spatially complete wind fields that are consistent with the available sparse observations. Unlike methods that directly correct ERA5 or WRF background fields, KiloDA does not use a contemporaneous model field as an initial state. Rather than performing conventional station interpolation, it reconstructs wind fields through observation-constrained posterior sampling.

### Learning diffusion priors from WRF fields

The generative component of KiloDA is based on a denoising diffusion probabilistic model (DDPM)[29]. Let $x_0$ denote a clean WRF wind-field sample. The forward diffusion process progressively perturbs $x_0$ with Gaussian noise over $T$ diffusion steps. Defining $\alpha_t = 1 - \beta_t$ and $\bar{\alpha}_t = \prod_{s=1}^{t} \alpha_s$, a noisy sample at diffusion step $t$ is written as

$$x_t = \sqrt{\bar{\alpha}_t} x_0 + \sqrt{1 - \bar{\alpha}_t}\epsilon, \quad \epsilon \sim \mathcal{N}(0, I), \tag{3}$$

where $\beta_t$ is specified by the noise schedule. The reverse process progressively removes noise from $x_t$ and is parameterized as

$$p_\theta(x_{t-1} \mid x_t) = \mathcal{N}\big(x_{t-1}; \mu_\theta(x_t, t), \Sigma_\theta(x_t, t)\big), \tag{4}$$

where $\mu_\theta$ and $\Sigma_\theta$ denote the parameterized mean and diagonal covariance of the reverse transition. A U-Net predicts the noise component $\epsilon_\theta(x_t, t)$ in the current noisy state. The noise-prediction objective is

$$\mathcal{L}_{simple} = \mathbb{E}_{x_0,t,\epsilon}[||\epsilon - \epsilon_\theta(x_t, t)||_2^2]. \tag{5}$$

This objective trains the noise-prediction component of the reverse denoising process. After training, sampling starts from Gaussian noise $x_T \sim \mathcal{N}(0, I)$ and progressively applies the learned reverse process to generate kilometer-scale near-surface wind fields that follow the distribution of the WRF training data. Neither station observations nor ERA5 fields were provided directly to the diffusion models during training; the models therefore learned priors for spatially complete wind fields from the WRF archive.

Separate diffusion models were trained for U10 and V10 using hourly WRF data from 2024. Each model used a U-Net with 64 base channels and channel multipliers of (1, 2, 3, 4). The reverse-process variance was parameterized using the learned-range formulation[39]. Training used AdamW with a learning rate of $1 \times 10^{-4}$ and a global batch size of 32. The diffusion process comprised 1,000 steps with a linear noise schedule from $\beta_1 = 1 \times 10^{-4}$ to $\beta_T = 2 \times 10^{-2}$. The exponential moving average (EMA) checkpoint saved at 700,000 optimization steps was used for reconstruction, with a full 1,000-step reverse process.

## Posterior sampling constrained by station observations

A pretrained diffusion model can generate kilometer-scale wind-field samples from Gaussian noise, but reconstruction of the target wind field must also be consistent with the available station observations. KiloDA therefore introduces an observation-consistency constraint during reverse diffusion, transforming wind-field generation into an observation-constrained posterior sampling problem[40]. Let $x_0$ denote the complete gridded wind field, $y$ the sparse station observations represented on the model grid and $A$ the observation operator. The observation model is

$$y = A(x_0) + \eta, \tag{6}$$

where $\eta$ represents observational error. The observation operator $A$ is implemented using the station-observation mask $M$ as

$$A(x) = M \odot x, \tag{7}$$

where $M$ is a binary station-observation mask with $M = 1$ at observed grid cells and $M = 0$ elsewhere, and $\odot$ denotes element-wise multiplication. At each reverse-diffusion step, the U-Net predicts the noise component $\epsilon_\theta(x_t, t)$ from the current noisy state $x_t$, from which the clean wind-field estimate is obtained as

$$\hat{x}_0 = \frac{1}{\sqrt{\bar{\alpha}_t}}\left[x_t - \sqrt{1 - \bar{\alpha}_t}\epsilon_\theta(x_t, t)\right]. \tag{8}$$

KiloDA then evaluates the discrepancy between the estimated and observed values at the constrained locations. The corresponding observation-consistency gradient is expressed as

$$\nabla_{x_t} \|y - A(\hat{x}_0)\|_2^2. \quad (9)$$

At each reverse-diffusion step, this gradient is applied to the sampling update to improve consistency with the station observations. The guidance scale controls the relative contributions of observational consistency and the learned diffusion prior. A larger guidance scale places greater weight on observational consistency, whereas a smaller value gives greater weight to the diffusion prior. In this study, a constant guidance scale of 6 was used at all reverse-diffusion steps.

For each reconstruction time, KiloDA sampled the two wind components separately, generating 10 ensemble members for each. Corresponding U10 and V10 members were then paired to calculate WS10, from which the ensemble mean was obtained. The set of stations used for observational conditioning was varied between experiments by changing the observation mask $M$.

## Experimental design

### Idealized WRF experiments

Idealized experiments based entirely on WRF fields were used to evaluate the ability of KiloDA to recover a complete kilometer-scale wind field from sparse observations under controlled conditions in which the full reference field was known. Independent WRF fields that were not used for model training served as the reference fields. Pseudo-observations were extracted from selected grid points and supplied as the only constraints to KiloDA and the interpolation baselines.

Two experimental configurations were used. In the station-like experiment, pseudo-observations were extracted from WRF grid cells corresponding to the actual station locations, and reconstruction performance was evaluated over JFM 2025. In the random-grid experiment, 100–4,000 grid points were randomly sampled as pseudo-observations to examine the sensitivity of reconstruction skill to the number of constraints during a 32-hour strong wind period. Reconstruction performance in both experimental configurations was further examined across terrain relief and elevation classes to assess robustness under different terrain conditions. Elevation was divided into low (<800 m), moderate (800–1,500 m) and high (≥1,500 m) classes. Terrain relief was defined as the difference between the maximum and minimum terrain elevations within a 3 × 3 grid-cell neighborhood and was divided into low (<100 m), moderate (100–300 m) and high (≥300 m) classes. These thresholds were selected with reference to the spatial distributions of elevation and terrain relief across the study domain (Supplementary Fig. 2) to distinguish relatively low and flat terrain from high-elevation and strongly varying terrain.

Non-station grid cells in the station-like experiment and unconstrained grid cells in the random-grid experiment were used for independent verification.

### Real-observation validation experiments

A contiguous withheld-region validation was used to evaluate whether KiloDA could reconstruct wind fields within a contiguous region with no internal observational constraints using only observations outside the region. The target region was identified from the dominant spatial cluster of stations whose combined ERA5 wind-component RMSE relative to observations ranked in the top 20% during JFM 2025, and an expanded rectangular region enclosing this cluster was defined as the withheld region. All 238 stations within the target region were withheld for independent evaluation, while the remaining 1,102 stations outside the region were used as observational constraints. This design avoids potentially optimistic evaluation at constrained stations and directly assesses reconstruction in the withheld region.

As a complementary robustness test, 100 stations were randomly withheld from complex-terrain locations with terrain relief ≥300 m or elevation ≥1,500 m and used only for independent evaluation, while the remaining 1,240 stations provided observational constraints. This experiment assessed the robustness of reconstruction performance under a spatially dispersed holdout configuration.

### Regional influence experiments

Regional influence experiments were conducted during the representative 32-hour strong wind period to examine how the remote reconstruction performance of KiloDA varied with the spatial location of observational constraints. Stations within the study domain were divided into nine subregions (R0–R8), each containing approximately the same number of stations. In each experiment, observations from only one subregion were used to constrain KiloDA, while reconstruction performance was evaluated separately in each of the remaining eight subregions, which served as target regions.

For each combination of source and target regions, the remote impact (RI) was quantified as the percentage reduction in WS10 RMSE in the target region achieved by KiloDA relative to ERA5:

$$RI_{s,t} = \frac{RMSE_{ERA5,t} - RMSE_{KiloDA,s,t}}{RMSE_{ERA5,t}} \times 100\% \tag{10}$$

Mean RI was obtained by averaging across the eight corresponding target regions for each source region and across the eight source regions for each target region. Positive RI values indicate lower target-region RMSE for KiloDA than for ERA5, whereas negative values indicate higher RMSE.

### Statistics and reproducibility

Reconstruction accuracy was quantified using RMSE, spatial PCC and TCC, with PCC calculated across spatial locations and TCC over time at each withheld station. PDFs and radially averaged PSDs were used to assess distributional and spatial-scale characteristics. No inferential hypothesis tests were performed. The station-like and random-grid evaluations used 2,117 and 32 hourly WRF reference fields, respectively, while the contiguous-region and complex-terrain random holdouts included 238 and 100 withheld stations. Ensemble members were used only to obtain the ensemble-mean WS10 field and were not treated as independent statistical replicates. Random selections used a fixed seed of 0. Random-grid constraint masks were independently sampled for each constraint number and fixed across evaluation times, while validation and assimilation masks in the random-holdout experiment were fixed throughout. The same selected locations were used for all compared methods where applicable.

## Data availability

The data used in this study include WRF near-surface wind simulations, surface meteorological station observations and ERA5 reanalysis data. ERA5 hourly 10-m wind component data are available from the Copernicus Climate Data Store (https://cds.climate.copernicus.eu/datasets/reanalysis-era5-single-levels). The WRF simulations were generated by the authors and were used to train the kilometer-scale diffusion priors in KiloDA. Owing to the large data volume, the full WRF training archive is not publicly available but is available from the corresponding author upon reasonable request. The raw surface-station observations used for observational conditioning and evaluation are subject to third-party data-use agreements and cannot be publicly redistributed. Processed data supporting the analyses and figures are available from the corresponding author upon reasonable request.

## Code availability

The custom code supporting the findings of this study is available from the corresponding author upon reasonable request during peer review and will be made publicly available in a permanent repository upon publication.

## Funding

This work was supported by the Smart-Grid National Science and Technology Major Project (Grant No. 2025ZD0805500).

## Acknowledgements

The authors thank the China Meteorological Administration for providing the surface meteorological station observations and the European Centre for Medium-Range Weather Forecasts for providing the ERA5 reanalysis data. The authors acknowledge the technical support provided by the National Large Scientific and Technological Infrastructure "Earth System Numerical Simulation Facility" (https://cstr.cn/31134.02.EL).

## Author contributions

Ya Wang conceived the study and designed the experiments. Yihan Zheng developed the KiloDA framework, performed the model experiments, conducted the data analysis, prepared the figures, and wrote the initial draft of the manuscript. Ya Wang and Gang Huang supervised the project and provided primary intellectual guidance. Shenming Fu contributed to the interpretation and discussion of the results and the revision of the manuscript. Haijie Li assisted with troubleshooting issues related to model training and inference. Yujiang Cai contributed to discussions of the data analysis and interpretation of the results. All authors discussed the results and commented on the final version of the manuscript.

## Competing interests

The authors declare no competing interests.

# Supplementary Information for

# KiloDA: Reconstructing kilometer-scale near-surface wind states from sparse station observations

Yihan Zheng[1,3], Ya Wang*[2,3], Gang Huang*[2,3], Shenming Fu[2,3], Haijie Li[2,3] and Yujiang Cai[2,3]

[1] *Center for Monsoon System Research, Institute of Atmospheric Physics, Chinese Academy of Sciences, Beijing 100029, China.*

[2] *China Key Laboratory of Earth System Numerical Modeling and Application, Institute of Atmospheric Physics, Chinese Academy of Sciences, Beijing 100029, China.*

[3] *University of Chinese Academy of Sciences, Beijing 100049, China.*

**Contents of this file**

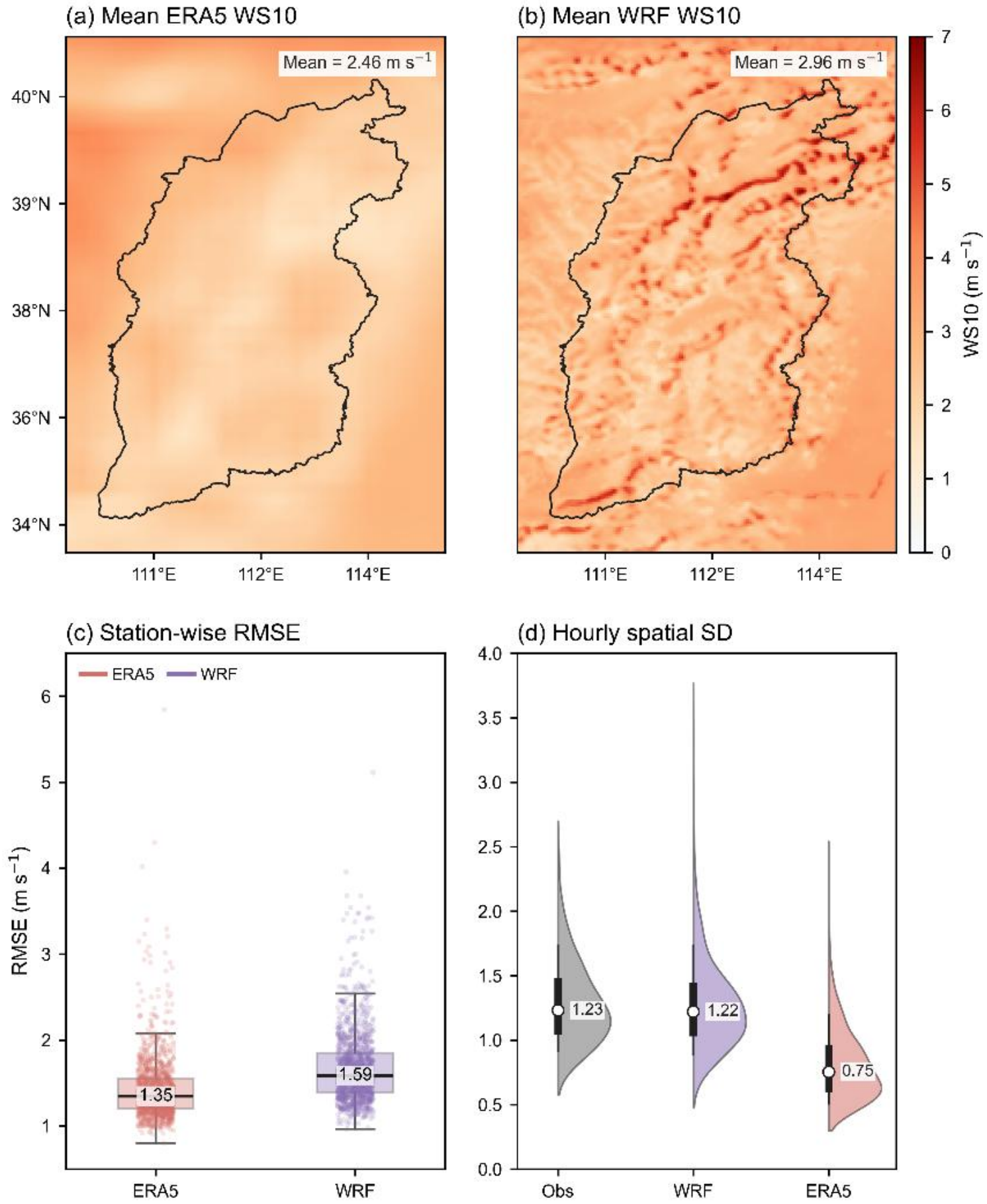

**Supplementary Fig. 1 Comparison of near-surface wind characteristics among the Weather Research and Forecasting (WRF) model, ERA5 atmospheric reanalysis and station observations in 2024.** (a, b) Mean 10-m wind speed (WS10) fields from ERA5 and WRF, respectively. (c) Station-wise WS10 root mean square error (RMSE) of ERA5 and WRF against surface-station observations. (d) Distributions of hourly spatial standard deviation (SD) of WS10 for observations, WRF and ERA5 at common valid station locations. WS10, RMSE and SD are expressed in m $s^{-1}$.

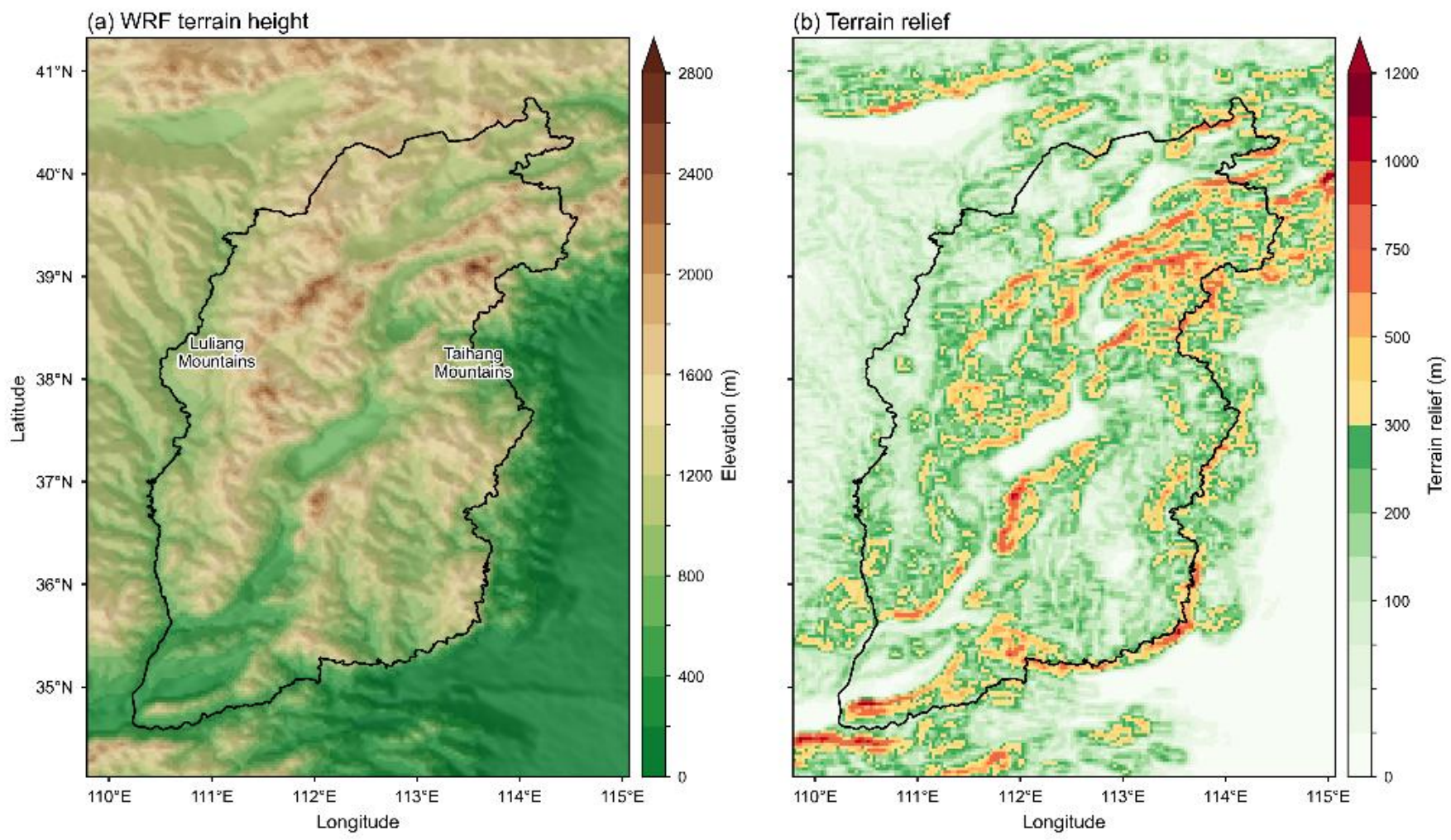

**Supplementary Fig. 2 Terrain characteristics of the study domain.** (a) WRF terrain height over the study domain, with the administrative boundary shown in black. The Lüliang and Taihang Mountains are labeled. (b) Terrain relief, defined as the difference between the maximum and minimum terrain elevations within each 3 × 3 grid-cell neighborhood. Terrain height and terrain relief are in m.

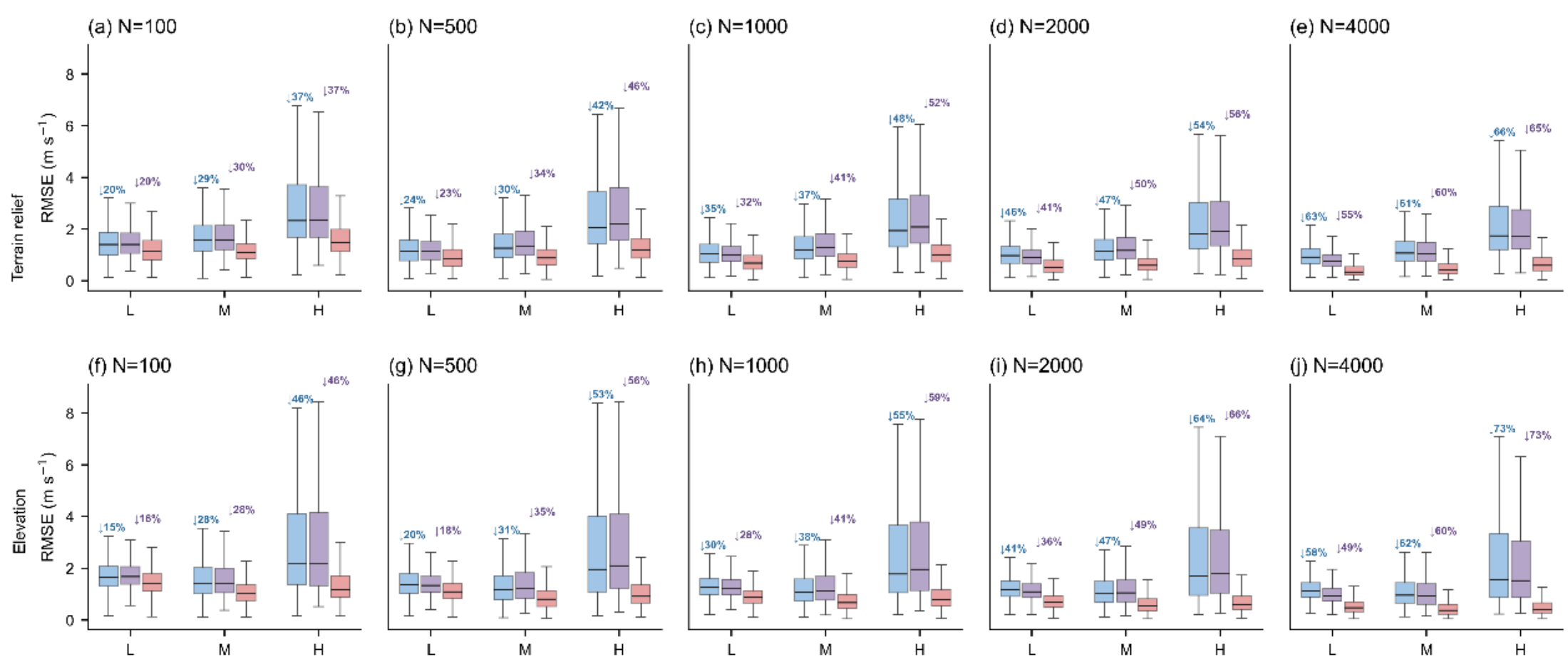

**Supplementary Fig. 3 Sensitivity of WS10 reconstruction error to the number of constraint grid cells and terrain conditions in the random-grid experiment.** Distributions of WS10 RMSE for inverse distance weighting (IDW), Kriging and KiloDA across low (L), moderate (M) and high (H) terrain-complexity classes, using N = 100, 500, 1,000, 2,000 and 4,000 randomly selected constraint grid cells during the representative 32-hour strong-wind period. Panels (a–e) stratify results by terrain relief, whereas panels (f–j) stratify results by elevation. Boxes show the interquartile range, center lines indicate medians, and whiskers extend to the most extreme values within 1.5 times the interquartile range; outliers are not shown. Blue and purple percentage labels indicate the relative RMSE reduction of KiloDA compared with IDW and Kriging, respectively. RMSE is in m $s^{-1}$.

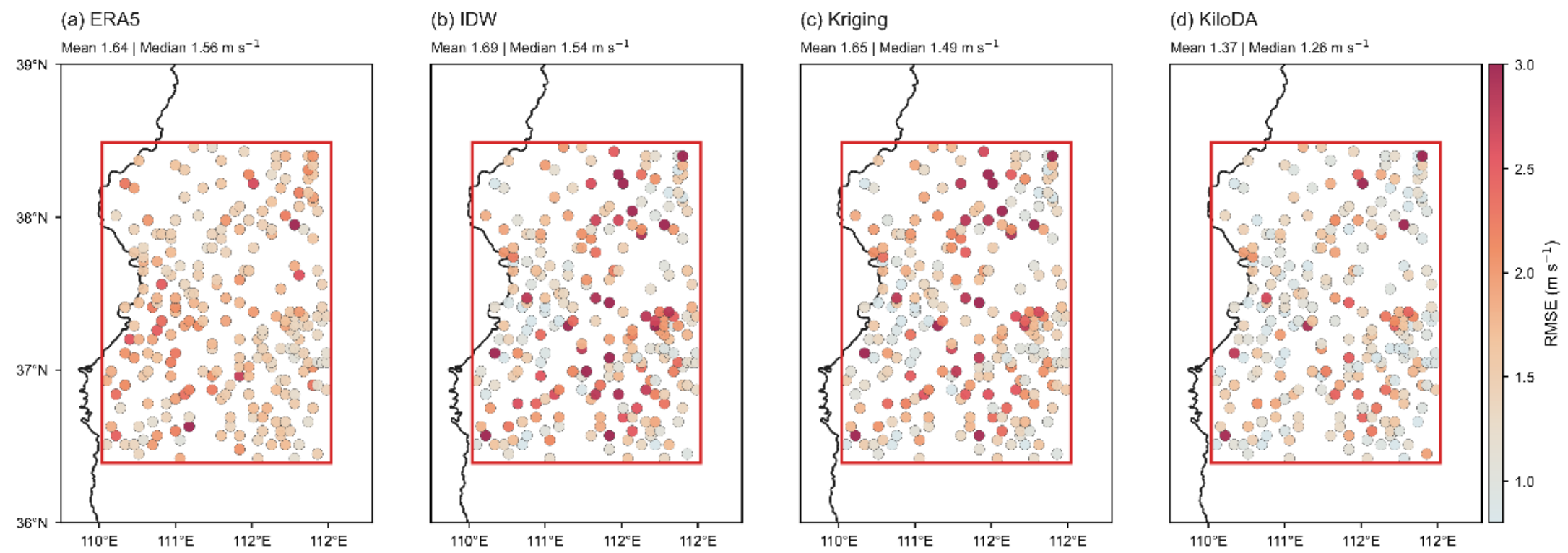

**Supplementary Fig. 4 Spatial distributions of station-wise WS10 errors in the contiguous withheld region.** Station-wise WS10 RMSE for (a) ERA5, (b) IDW, (c) Kriging and (d) KiloDA during January to March 2025. Each marker denotes a withheld station and is colored by its RMSE. All panels use the same color scale; panel subtitles report the mean and median station-wise RMSE. The red rectangle marks the withheld target region. RMSE is in m $s^{-1}$.

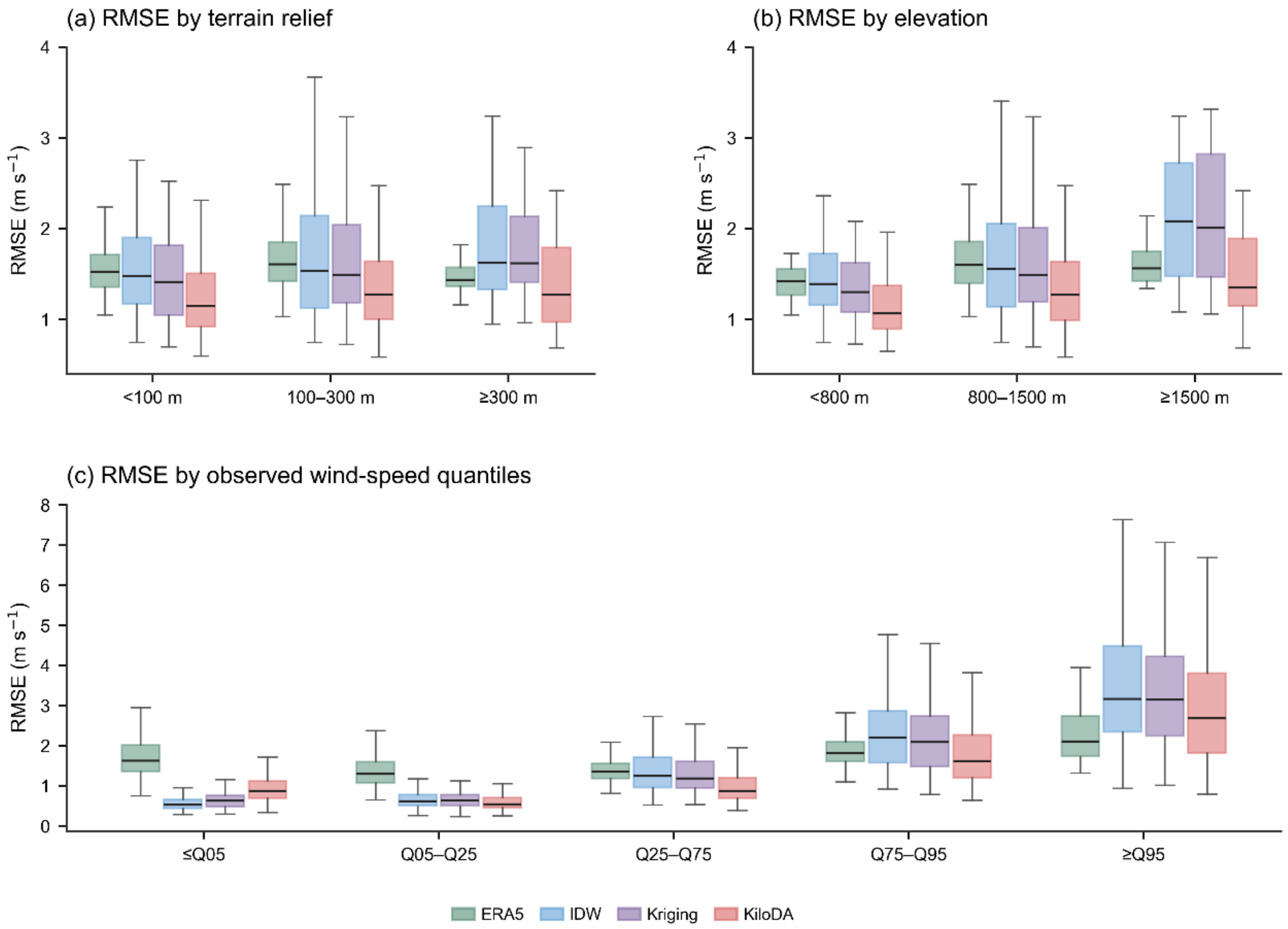

**Supplementary Fig. 5 Stratified WS10 reconstruction errors for the contiguous withheld-region validation.** Distributions of station-wise WS10 RMSE for ERA5, IDW, Kriging and KiloDA at the withheld stations, stratified by (a) terrain relief, (b) station elevation and (c) observed wind-speed quantiles during January to March 2025. Boxes show the interquartile range, center lines indicate medians, and whiskers extend to the most extreme values within 1.5 times the interquartile range; outliers are not shown. RMSE is in m $s^{-1}$.

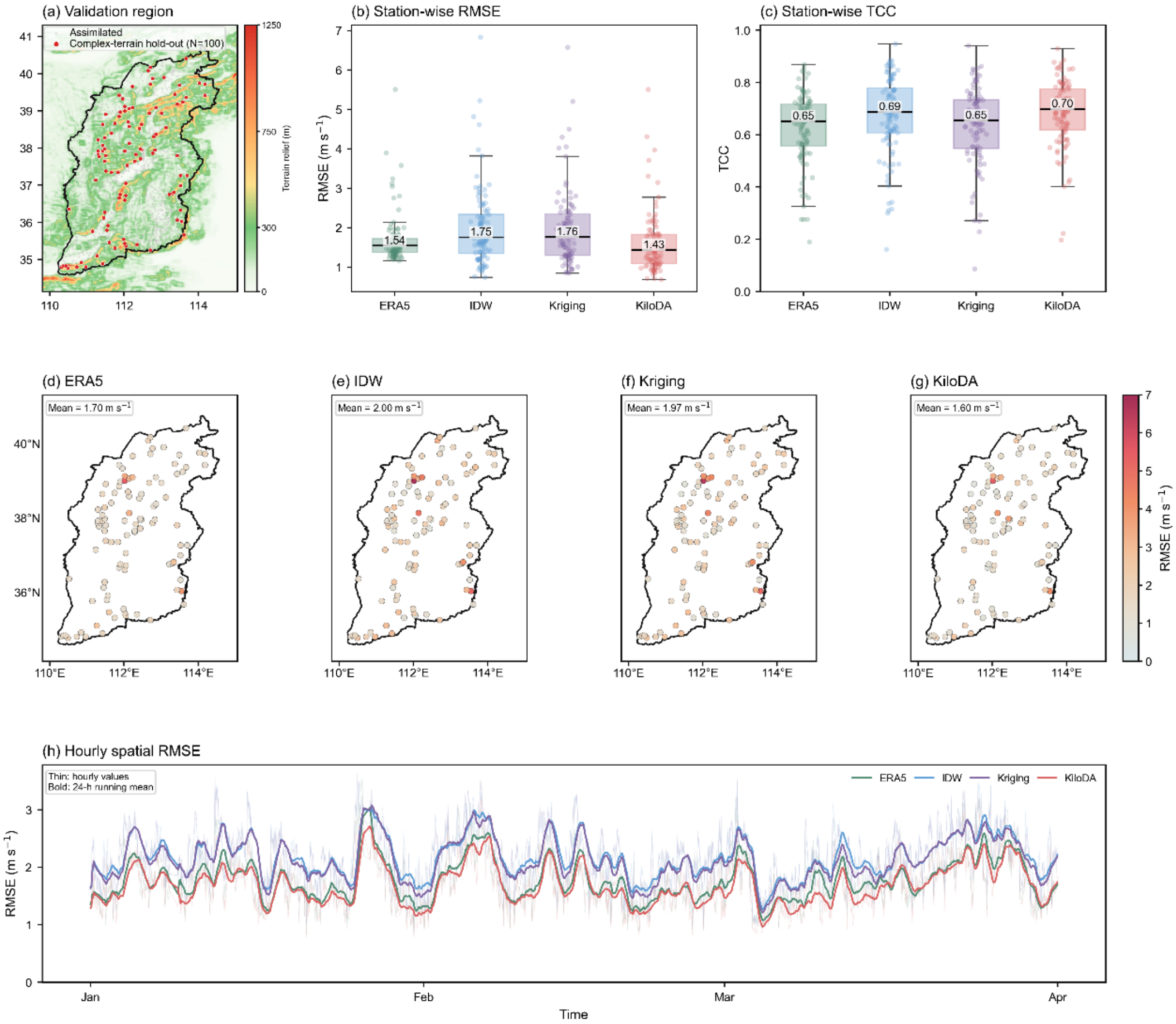

**Supplementary Fig. 6 Random holdout validation over complex terrain.** Results are shown for the January to March 2025 evaluation period using a random holdout of 100 complex-terrain stations. (a) Terrain-relief map of the study domain and locations of the 100 randomly withheld stations. (b, c) Distributions of station-wise WS10 RMSE and temporal correlation coefficient (TCC), respectively, for ERA5, IDW, Kriging and KiloDA. Boxes show the interquartile range, center lines indicate medians, whiskers extend to the most extreme values within 1.5 times the interquartile range, and points denote individual withheld stations; numbers indicate median values. (d–g) Spatial distributions of station-wise RMSE for ERA5, IDW, Kriging and KiloDA, respectively; inset labels show the mean RMSE. (h) Hourly spatial RMSE across the 100 randomly withheld stations, with thin lines showing hourly values and bold lines showing trailing 24-h means.